\documentclass[preprint,aps,amsmath,superscriptaddress,nofootinbib,tightenlines]{revtex4}
\usepackage{epsfig}

\begin{document}

\title{
\vspace{10pt}
\huge{
An Eta Primer:\\ 
Solving the $U(1)$ Problem with AdS/QCD}
\vspace{10pt}
}

\author{Emanuel Katz}
\affiliation{Department of Physics, Boston University, Boston, MA 02215, USA}

\author{Matthew D.~Schwartz}
\affiliation{Department of Physics, Johns Hopkins University, Baltimore, MD 21218, USA
\vspace{15pt}}

\begin{abstract}
Inspired by the AdS/CFT correspondence, 
we study the pseudoscalar mesons of QCD through a dual embedding
in a strongly curved extra dimensional spacetime.
This model incorporates the consequences of symmetry and has very few free parameters,
due to constraints from five-dimensions and the operator product expansion of
QCD. Using as inputs $f_\pi$ and the pion, kaon, and rho masses, we
compute the eta and eta prime masses to be 520 and 867 MeV, respectively.
Their decay rates into photons are also computed and found to be in 
good agreement with data.
\thispagestyle{empty}
\end{abstract}

\maketitle
\newpage

\section{Introduction}

While QCD has been unequivocally established as the theory of strong
interactions, the resolution of one of its mysteries, the $U (1)$ problem, has
remained somewhat unsatisfying. The $U (1)$ problem is that the Lagrangian of
QCD
\begin{equation}
  \mathcal{L}_{\mathrm{QCD}} = \frac{1}{4 g_s^2} G_{\mu \nu}^2 + \bar{q}^i D \!\!\!\! \slash\: q^i + M_{q_i}
  \bar{q}_i q_i
\end{equation}
has, in the massless limit, a global chiral $U (1)_A$ symmetry, under which
$q_i \rightarrow e^{i \theta \gamma_5} q_i$,
which does not seem to be
reflected in the spectrum of light pseudoscalar mesons. The formation of quark
condensates $\langle \bar{q}_i q_j \rangle \approx \Lambda_{\mathrm{QCD}}^3
\delta_{i j}$ spontaneously beaks the $U(3)_L \times U (3)_R$ symmetry of
massless QCD down to a diagonal $U(3)_V$, which should result in nine
pseudoscalar pseudogoldstone bosons. The problem is that, with masses
included, chiral perturbation theory unambiguously predicts a neutral pseudoscalar
meson whose mass is strictly less than $\sqrt{3} m_{\pi}$~\cite{Weinberg:1975ui}. 
However, the true
hadron spectrum contains only the regular $\pi^0$ (140), the $\eta$ (549),
and the $\eta'$(957), so the chiral perturbation theory bound is
clearly violated.

Actually, the $U (1)$ problem is little more subtle; the $U (1)_A$ is
anomalous, {\it{i.e.}} broken by quantum effects. Mathematically, while the QCD
Lagrangian is invariant (except for the mass terms), the functional measure in
the path integral is not, and so a chiral rotation results in
\begin{equation}
  \mathcal{L}_{\mathrm{QCD}} \rightarrow \mathcal{L}_{\mathrm{QCD}} + \theta
  \frac{\alpha_s}{8 \pi} G_{\mu \nu} \tilde{G}_{\mu \nu} \label{thetaeq}
\end{equation}
where $\tilde{G}_{\mu \nu} \equiv \frac{1}{2} \epsilon_{\mu \nu \alpha \beta}
G_{\alpha \beta}$. The simplest solution of the $U(1)$ problem is then to say
that the $U(1)_A$ symmetry is not really a symmetry at all, so there should
be no corresponding pseudogoldstone boson~\cite{'tHooft:1999jc,'tHooft:1986nc}.
However, from the QCD side it is hard to see how
the new term in \eqref{thetaeq} could make any difference. Because it is a total
derivative, any Feynman diagram involving the anomaly carries a factor of zero
for total momentum. Thus, the new term does not contribute at any order in
perturbation theory, and therefore the solution must be non-perturbative.

If we accept that $U(1)_A$ is not a symmetry, the pseudoscalar sector can be modeled
in the chiral Lagrangian~\cite{'tHooft:1999jc,Nath:1980nf,Kaiser:1998ds}. In full generality,
the chiral Lagrangian has five free parameters, corresponding to a scale for the $U(1)$ breaking,
and four decay constants characterizing the interaction strength between the $\eta$ and $\eta'$ mesons
and the $J_\mu^{(0)}$ and $J_\mu^{(8)}$
currents~\cite{Nath:1980nf,Kaiser:1998ds}. The number of parameters can be reduced, for example 
by going to the large $N_c$ limit, but a few parameters at least remain. In particular, none of the
assumptions allow a first-principles calculation of the $\eta'$ mass.

The first convincing resolution of the $U(1)$ problem was given 
by 't Hooft~\cite{'tHooft:1976fv},
who argued topological instanton configurations of the QCD vector potential
can contribute to the path integral through the anomaly. The instanton
contributions are suppressed by factors of $\exp (-1 / g_s^2)$, which
can be significant only for large $g_s$, {\it{i.e.}} when QCD enters the
non-perturbative regime. However, instanton calculations generically have
infrared divergences, due to integrals over large instanton size, so it is
impossible to use them for precise quantitative calculations. Nevertheless,
they seem to be the correct qualitative solution. And, for example, using QCD
sum rules~\cite{Novikov:1979ux}, they can be used to get a ballpark estimate of the $\eta'$ mass
($\sim 1$ GeV).

Other non-perturbative insights into the $U(1)$ problem have come from the
lattice~\cite{Aoki:2006xk,DelDebbio:2004mc}. 
Because the $\eta'$ is critically sensitive to both quark loops and
non-local field configurations, it has been a challenge to simulate.
Nevertheless, the lattice has been remarkably successful in this case, and a
recent estimate~\cite{Aoki:2006xk} puts the $\eta'$ at $871 \pm 46$ MeV, which is within 10\% of
the experimental value. This is absolute confirmation that QCD itself solves
the $U(1)$ problem. But it is hard to get any qualitative understanding from
such a purely numerical approach.

In this paper, we propose that the pseudoscalar mesons can be studied both
qualitatively and quantitatively with a non-perturbative framework based on the
AdS/CFT correspondence~\cite{Maldacena:1997re}. 
This framework has already produced an impressive
post-diction of the meson spectrum, decay constants,
and couplings~\cite{Erlich:2005qh,DaRold:2005zs,Katz:2005ir}. It has also led
to some new insights into observations about QCD, such as vector meson
dominance~\cite{DaRold:2005zs}, and the structure of tensor mesons~\cite{Katz:2005ir}.
Thus, it is natural to ask whether it
can say anything about the $U(1)$ mystery of QCD.

The approach to AdS/QCD we take in this paper is completely bottom up.
Although the AdS/CFT correspondence began strictly as a duality between a
four-dimensional conformal gauge theory and a 10-dimensional string theory, it
is difficult to make any quantitative predictions about QCD from the string
side. Early work, studying for example, the glueball spectrum of a large-$N$
theory~\cite{Csaki:1998qr}, had some success when compared to lattice results; but to study real
world QCD, with three flavors and massive quarks, we would need much more
information about the string dual of QCD than is currently known 
(see, for example~\cite{Sakai:2004cn}).
Instead, we
assume that whatever the string theory is, it must contain bulk modes
dual to the various local operators of QCD.  To reproduce the conformal behavior of the
asymptotically free regime of QCD, these modes will propagate on a
background close to Anti-deSitter space.  
It turns out this is enough
information to reproduce a number of non-trivial quantitative predictions
about QCD at low energy.

\section{Setup}
The setup is a five-dimensional space, with background metric
\begin{equation}
  d s^2 = \frac{w (z)^2}{z^2} ({\mathrm{d}}x^2_\mu - {\mathrm{d}}z^2)
\end{equation}
In pure AdS the warp factor is $w (z) = 1$, but we will allow for
background corrections due to deviations from conformality.
The extra dimension can be thought of as energy, with small $z$
representing high energy. Thus we model the IR, where QCD is strong, by boundary
conditions at a point $z_m \sim 1 / \Lambda_{\mathrm{QCD}}$. We also impose
boundary conditions at $z = 0$, high energy, where QCD approaches a trivial
conformal fixed point. Thus the gravity background is modeling energies between
$\Lambda_{\mathrm{QCD}}$ and infinity. 

For this 5-D description to be equivalent to QCD, it should reproduce QCD
correlation functions of external currents. These appear as probes in the UV.
For each QCD operator which couples the vacuum to these currents, there must
be a corresponding field in 5D which also couples to the current. More
generally, for each operator in QCD, there should be a 5D field. In our case,
the currents of interest are $J_{\mu R}^b$ and $J_{\mu L}^b$, the right- and
left-handed $U (3)$ currents. The corresponding fields are bulk gauge fields
$A_L$ and $A_R$. The operator $\bar{q}_i q_j$ which spontaneously breaks $U
(3) \times U (3) \rightarrow U (3)_V$ is represented by bifundamental 
bulk scalars $X_{i j}$. The fields X$_{i j}$ have interactions and a
potential. However, this potential is neither calculable nor relevant to low
energy, so we simply parameterize this potential and fit to data.

To study the $U(1)$ problem, we now introduce a new complex field $Y$ to represent 
the square of the gluon field
strength: $Y \sim G^2_{\mu \nu}$. We can think of the phase of $Y$
as dual to $G \tilde{G}$. We emphasize that identifying $Y$ is not
important for the low energy physics, we only use it to manifest a linear
representation of the symmetries. Thus our 5D Lagrangian, including
all terms allowed by symmetry, is
\begin{equation}
  \mathcal{L} = \sqrt{g} \left\{ - \frac{1}{4 g_5^2} (F_L^2 + F_R^2) +
  {\mathrm{Tr}} \left\{{|D X|^2 + 3 |X|^2}\right\} 
+ \frac{1}{2}|D Y|^2  + \frac{\kappa}{2}[Y^{N_f} \det(X) + {\mathrm{h.c.}}] \right\} \label{linearsig}
\end{equation}
That $X$ gets a 5D mass but $Y$ does not follows form the AdS/CFT map between
masses and dimensions of operators. With these masses, the solutions to the equations of motion for $X$
and $Y$ in pure $\mathrm{AdS}_5$ are
\begin{eqnarray}
  \langle X_{i j} \rangle &=& v_{i j}(z) \equiv \sigma_{i j} z^3 + m_{i j} z \label{xvev}\\
  \hspace{1em} \langle Y \rangle &=& \Xi z^4 + C \label{yvev}
\end{eqnarray}
These must correspond to the vacuum expectation values, $\langle
\bar{q}_i q_j \rangle \sim \sigma_{i j} \sim \Lambda_{\mathrm{QCD}}^3$ and
$\langle G_{\mu \nu}^2 \rangle \sim \Xi \sim \Lambda_{\mathrm{QCD}}^4$ and to the
sources $M_q$ and $g_s$. Thus the $z$-dependence of a field is seen to match
the scaling dimension of the corresponding operator. In the 3-flavor
case, for simplicity, we will assume that $\sigma_{i j} = \sigma \delta_{i j}$
({\it{i.e.}}, $\langle \bar{s} s \rangle = \langle \bar{d} d \rangle =
\langle \bar{u} u \rangle)$ and use only two masses $\hat{m} = \frac{1}{2}
(m_u + m_d)$ and $m_s$.  

To study this theory, we will explore the pseudoscalar excitations around the
$X$ and $Y$ backgrounds.
\begin{eqnarray}
X_{i j} &=& \langle X_{i j} \rangle \exp (i \eta^b \tau^b)\\
Y &=& \langle Y \rangle \exp(i a/\sqrt{2N_f})
\end{eqnarray}
There are of course scalar excitations as
well, but these are harder to study as they are sensitive to details of the
$X$ and $Y$ effective potentials. For the left and right $U (3)$ gauge fields,
we will only need the axial combination $A = A_R - A_L$.  
The longitudinal modes of $A_\mu$ mix with the pions, so it is helpful to
include them explicitly with the replacement $A_{\mu} \to \partial_{\mu} \varphi$.
Then we get
\begin{eqnarray}
  \mathcal{L} &=& 
\frac{1}{2 g_5^2 z} (\partial_{\mu} A_5^b - \partial_z  \partial_{\mu} \varphi^b)^2 
+ \sum_{\mathrm{flavors}} 
\frac{v^2}{2 z^3}[(\partial_\mu \varphi^b-\partial_\mu\eta^b)^2-(A_5^b  -
\partial_z \eta^b)^2] \nonumber \\
&& 
+\frac{C^2}{2 z^3}[(\partial_\mu \varphi^0-\partial_\mu a)^2-(A_5^0  - \partial_z a)^2]
+  \frac{\kappa}{2 z^5} v^{N_f} (a - \eta^0)^2 \label{pieq}
\end{eqnarray}
The $0$ on $\eta^0$
refers to the
 $\tau^0 = \frac{1}{\sqrt{6}} {\mathrm{diag}}(1,1,1)$ generator of $U(3)$, and an $8$
superscript will refer to the $\tau^8 = \frac{1}{\sqrt{12}} {\mathrm{diag}}(1,1,-2)$ generator,
in the $u, d, s$ basis.
Note that
we have absorbed a constant into the definition of $C$,
and absorbed factors of $C$ into the definition of $\kappa$. 
We have also dropped $\Xi \sim \langle G_{\mu \nu}^2 \rangle$, as it is will be a
subleading power correction in everything that follows.
Regardless of these conventions, it is simplest to regard Eq.~\eqref{pieq}, 
instead of Eq.~\eqref{linearsig},
as the starting point for phenomenological analysis.

In 4D the $U(1)_A$ symmetry is anomalous; it is broken by quantum effects.
But quantum effects in 4D correspond to classical effects in 5D, so the
symmetry should be explicitly broken in 5D. In unitary gauge, this is true. But
in the form \eqref{pieq}, we have restored the symmetry with our ``axion'' Goldstone
boson $a$. In fact the whole $U(3)_A$ is gauged, so there is a local symmetry
under which
\begin{eqnarray}
  A_M^b &\to& A_M^b + \partial_M \alpha^b \nonumber\\
  \eta^b &\to& \eta^b + \alpha^b \nonumber\\
  a     &\to& a + \alpha^0   \nonumber
\end{eqnarray}
We can use this to set $A_5^b = 0$, however it is helpful to
retain these modes to simplify the calculations.

Because of the gauge symmetry, the fields $\varphi, \eta$ and $a$ are not
strictly independent, but they do have different physical meanings as can be seen
by introducing external sources (and notation). We define
\begin{equation}
  J_{\mu}^b \equiv \sum_i \bar{q}_i \gamma_{\mu} \gamma_5 \tau^b_{i j} q_j 
\end{equation}
The $U (1)_A$ current is normalized as
\begin{equation}
  J_{\mu}^0 \equiv \frac{1}{\sqrt{2 N_f}} \sum_i \bar{q}_i \gamma_{\mu}
  \gamma_5 q_i \label{u1norm}
\end{equation}
A source $J_{\mu}^b A_{\mu}^b \delta (z)$ on the UV brane leads to $\varphi^b
\partial_{\mu} J_{\mu}^b \delta (z)$ after introducing $\varphi$ and
integrating by parts. So the source for $\varphi$ is
\begin{equation}
  J_{\varphi}^b \equiv \partial_{\mu} J_{\mu}^b
\end{equation}
Finally $\eta^b$ and $a$ by definition correspond to specific 4D fields, so we
have
\begin{equation}
  J_{\eta}^b \equiv g_\eta \bar{q}_i \gamma_5 \tau^b_{i j} q_j, \hspace{1em} 
J_a \equiv g_a \frac{\alpha_s}{8 \pi^2} G \tilde{G}
\end{equation}
Note that we use constants $g_\eta$ and $g_a$ to normalize $J_\eta$ and $J_a$,
while the normalization of $J_{\mu}$ is set by the interaction strength
$g_5$ in the Lagrangian.

These currents help us identify our pseudoscalar fields. We see that although
$\varphi, \eta$ and $a$ all mix they still have physical meanings: for a
particular mode, $a$ is the ``glueball'' component and $\eta$ and $\varphi$ are
the ``quark'' components of the corresponding mesonic wavefunction, with
$\varphi$ related to the longitudinal mode of the axial vector field. 

\subsection{Matching to QCD}
We will now calculate the parameters in our model by matching to the QCD
operator product expansion (OPE). Let us start immediately with the case of
interest, 3-flavors, massive quarks, and a physical $\eta'$. We will need to
make use of the anomaly equation
\begin{equation}
  J_\varphi^0= \partial_{\mu} J_{\mu}^0 = \sqrt{2 N_f} \frac{\alpha_s}{8 \pi} G \tilde{G} +
 \frac{1}{\sqrt{2 N_f}} \sum_{\mathrm{flavors}}
i M_{q_i} \bar{q}_i \gamma_5 q_i
\end{equation}
The anomaly shows up in the OPE~\cite{Shifman:1978bx,Reinders:1981ww}
\begin{equation}
  \langle J^0_{\varphi} J^0_{\varphi} \rangle = -\frac{N_f \alpha_s^2}{16 \pi^4} Q^4  \log Q^2 
+ \frac{3}{16\pi^2 N_f}\sum_{\mathrm{flavors}} M_q^2 Q^2 \log Q^2 
+ \cdots \label{opeanom}
\end{equation}

As explained in detail elsewhere~\cite{Erlich:2005qh,Katz:2005ir},
correlations functions are calculated in 5D
by solving the equations of motion in the presence of a source. For example,
this lets us deduce that $g_5 = 2 \pi$ in the case of interest, $N_C = 3$. For the
$\varphi^0$ correlator, which is relevant for the anomaly, we can write
\begin{equation}
  \langle J^0_{\varphi} J^0_{\varphi} \rangle = - \frac{Q^2}{g_5^2} \lim_{z
  \rightarrow 0} \frac{\partial_z \varphi^0 (z)}{z} \label{corr}
\end{equation}
Here, $\varphi^0 (z)$ is a bulk-to-boundary propagator, that is, a solution to
the equations of motion with $\varphi^0 (0) = 1$. For this calculation, chiral
symmetry breaking is irrelevant to leading order and so we can set $v = 0$.
The $\varphi^0$ and $A_5^0$ equations of motion then become
\begin{eqnarray}
  \partial_z \frac{1}{z} \partial_z \varphi^0 - g_5^2 \frac{C^2}{z^3}
  (\varphi^0 - a) &=& 0 \\
 g_5^2 C^2 \partial_z a - Q^2 z^2  \partial_z \varphi^0 &=& 0 \label{A5eqn}\
\end{eqnarray}
These are solved perturbatively near $z = 0$ by
\begin{eqnarray}
  \varphi^0 &=& 1 - g_5^2 \frac{C^2}{4} \log( Q^2 z^2) + g_5^2 \frac{C^2}{16} z^2
  Q^2 \log (Q^2 z^2) + \cdots\\
 a &=& - \frac{1}{4} Q^2 z^2 + \cdots
\end{eqnarray}
Matching \eqref{corr} to \eqref{opeanom} leads to
\begin{equation}
  C = \frac{\alpha_s}{2 \pi^2} \sqrt{2 N_f} \label{Cdef}
\end{equation}

Note that for this matching we have assumed that $\alpha_s$ 
is constant in the UV.  Of course, $\alpha_s$ runs with scale, and it is
therefore reasonable to assume that 
$\alpha_s$ would be a function of $z$ as the 1-loop QCD $\beta$
function. Hence, we should take
\begin{equation}                                                                                                              
  C = \sqrt{6}\frac{\alpha_s}{2 \pi^2}, \hspace{1em} \alpha_s =
  \frac{1}{\beta_0 \log                                         
  (\Lambda_{\mathrm{QCD}} z)}, \hspace{1em} \beta_0 = \frac{1}{2 \pi} (                                                       
  \frac{11}{3} N_C - \frac{2}{3} N_f) \label{Ceq}                                                                             
\end{equation}
where $\Lambda_{\mathrm{QCD}} \approx z_m^{- 1}$.
The fact that $\alpha_s$ varies slowly in the UV, and that $\partial_z \alpha_s \sim
\alpha_s^2$, makes the above matching correct to leading order in $\alpha_s$.

Instead of sourcing $\varphi^0$ with $\partial_{\mu} J_{\mu}^0 \neq 0$, we can
also consider pure gluodynamics and source $a$ by turning on $G \tilde{G}$. 
This will fix the normalization of $J_a$. The QCD correlation function
of interest is~\cite{Novikov:1979ux}
\begin{equation}
  \chi_t (Q) \equiv \langle ( \frac{\alpha_s}{8 \pi} G \tilde{G}) 
( \frac{\alpha_s}{8 \pi} G
  \tilde{G}) \rangle = -\frac{\alpha_s^2}{32 \pi^4} Q^4 \log Q^2
\end{equation}
which should match
\begin{equation}
  \chi_t (Q) = \frac{1}{g_a^2} \langle J_a J_a \rangle = \frac{C^2}{g_a^2}
  \lim_{z \rightarrow 0} \frac{a \partial_z a}{z^3}
\end{equation}
For a solution with $a (0) = 1$. Solving the equations of motion
perturbatively 
\begin{equation}
  a = 1 + \frac{1}{4} Q^2 z^2 - \frac{1}{32} Q^4 z^4 \log Q^2 z^2 + \cdots
\end{equation}
lets us deduce that
\begin{equation}
  g_a = 2 \pi^2 \frac{C}{\alpha_s}= \sqrt{2 N_f} \label{gadef}
\end{equation}
We will use this later on to compute the topological susceptibility $\chi_t
(0)$.

Next, we would also like to consider modifications to the background due to
deviations from conformality, in particular the effect of the strange quark
mass. Consider the transverse part of the axial vector OPE
\begin{equation}
  \langle J^0_{\mu} J^0_{\nu} \rangle = (Q_{\mu} Q_{\nu} - \eta_{\mu \nu} Q^2)
  \Pi_A + \cdots
\end{equation}
\begin{equation}
  \Pi_A = - \frac{1}{8 \pi^2} \log Q^2 - \frac{1}{Q^4} M_q \langle \bar{q}
  q \rangle + \cdots \label{piv}
\end{equation}
The $\log Q^2$ term on the right hand side is determined by conformal invariance and is used to
fit $g_5$. The second term is the power correction in which we are interested
now. By dimensional analysis, this should modify the warp factor to
\begin{equation}
  w (z) = 1 + c_4 z^4, \hspace{1em} c_4 \sim M_q \langle \bar{q} q \rangle
\end{equation}
This warp factor modifies the axial-vector equation of motion to
\begin{equation}
  \partial_z \frac{1 + c_4 z^4}{z} \partial_z A - Q^2 \frac{(1 + c_4 z^4)}{z} A
  = g_5^2 \frac{(m_q z + \sigma z^3)^2}{z^3} A 
\end{equation}
We can solve this perturbatively in $c_4 \sim m_q \sigma$. For $m_q = \sigma= 0$, 
the solution is $A^0(z)=Q z \mathcal{K}_1(Q z)$. 
Then, using the AdS Green's function $K'(z,z')$~\cite{Randall:2001gb}, the perturbative inhomogeneous solution can 
be written as
\begin{eqnarray}
  A^1 (z) &=& \int_0^{\infty} d z' K (z, z')\left[ 2 g_5^2 m_q \sigma Q z^2 \mathcal{K}_1(Q z) + 4 c_4 Q^2 z^3 \mathcal{K}_0(Q z)\right] \\
 &=& \left( \frac{2 g_5^2 m_q \sigma}{3} + \frac{4 c_4}{3}\right)\frac{z^2}{Q^2} + {\mathcal O}(z^3)
\end{eqnarray}
Thus AdS gives
\begin{equation}
  \Pi_A = \lim_{z \rightarrow 0} \frac{1}{g_5^2 Q^2} \frac{\partial_z A}{z} =
  \frac{2}{g_5^2} \log Q^2 -
\left( \frac{4 m_q \sigma}{3} + \frac{8 c_4}{3 g_5^2}\right)\frac{1}{Q^4} + \cdots
\end{equation}
Comparing to \eqref{piv}, we deduce $g_5 = 2 \pi$ and
\begin{equation}
  c_4 = \frac{3}{8} g_5^2 (M_q \langle \bar{q} q \rangle - \frac{4}{3} m_q \sigma) 
= -\frac{\pi^2}{2} m_q \sigma
\end{equation}
The last equality follows from the Gell-mann-Oaks-Renner relation~\cite{Erlich:2005qh}.

\subsection{Fitting to data}
Having determined $g_5, C, g_a$ and $c_4$ from matching to the OPE, our
Lagrangian is complete. The remaining unknowns must be fit to data. We use
\begin{eqnarray}
  m_{\rho} = 770 \: \mathrm{MeV} \hspace{1em} &\Rightarrow& 
\hspace{1em} z_m^{- 1} =  323\: \mathrm{MeV} = \Lambda_{\mathrm{QCD}} \\
  f_{\pi} = 93\: \mathrm{MeV} \hspace{1em} &\Rightarrow& \hspace{1em}
 \sigma = (333 \: \mathrm{MeV})^3 \\
  m_{\pi} = 140 \: \mathrm{MeV} \hspace{1em} &\Rightarrow& 
\hspace{1em} \hat{m} =  2.22 \: \mathrm{MeV} \\
  m_K = 494\: \mathrm{MeV} \hspace{1em} &\Rightarrow& 
\hspace{1em} m_s = 40.0\:  \mathrm{MeV} 
\end{eqnarray}
This gives $c_4 = - 0.676$ for strange and $c_4 = -0.037$ for up/down. Thus we are
justified in only turning on $c_4$ for the strange quark. Keep in mind that 
although these``quark masses'' may seem small, care must be taken when comparing them to 
masses deduced from another scheme.

There is one more parameter in our Lagrangian that remains, $\kappa$. The
$\kappa$ term corresponds to an entirely non-perturbative effect. However, it
multiples a function which grows like $z^{3 N_f - 5}$, so we expect it to act
effectively like a boundary condition forcing $a (z_m) = \eta^0 (z_m)$. Thus, we
leave $\kappa$ as a free parameter and show that for the $\eta'$ the results
are fairly independent of $\kappa$ for large $\kappa$.

\section{The $\eta'$}
\begin{figure}[t]
  \begin{minipage}[]{.45\textwidth}
    \begin{center}  
      \epsfig{file=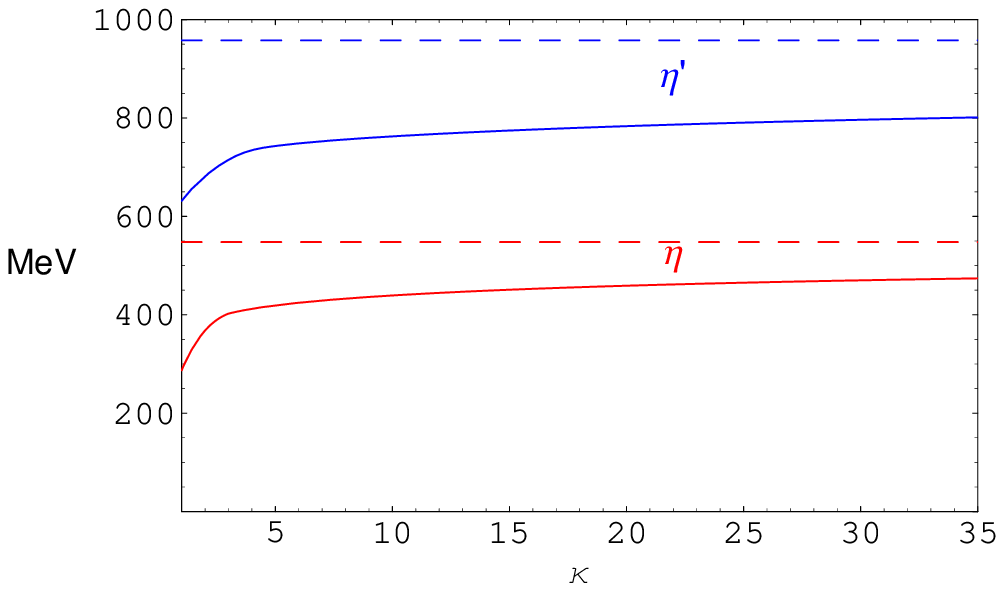, scale=0.75}
    \end{center}
  \end{minipage}
  \hfill
  \begin{minipage}[]{.45\textwidth}
    \begin{center}  
      \epsfig{file=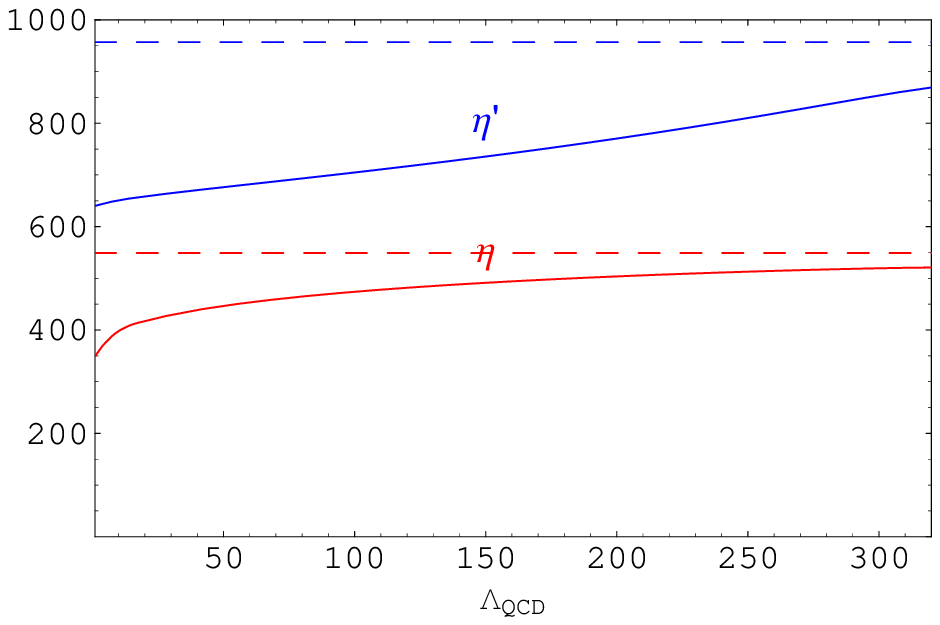, scale=0.75}
    \end{center}
  \end{minipage}
\label{figmkap}
\caption{Masses of $\eta$ and $\eta'$ as the parameters
 $\kappa$ and $\Lambda_{\mathrm{QCD}}$ are varied.
In the $\kappa$ plot, we fix $\Lambda_{\mathrm{QCD}} = 320$ MeV $\approx z_m^{-1}$ and in
the $\Lambda_{\mathrm{QCD}}$ plot, we take $\kappa = \infty$. Dashed lines are the experimental
values.}
\end{figure}
Having fit all the parameters in our Lagrangian (using only $m_{\rho},
f_{\pi}, m_{\pi}$ and $m_K$), we can now look at what masses are predicted. 
For the neutral pseudoscalar spectrum, there is a competition between
the quark masses, which force the $\eta$ and $\eta'$ into the $q$ and $s$
bases, and the anomaly and $\kappa$ terms, which push towards the $\eta^0$ and
$\eta^8$ basis. There are seven fields, $\varphi^{0, 8}, \eta^{0, 8}, A_5^{0,
8}$ and $a$, but we can use the residual gauge invariance to set $A_5^{0,8} = 0$.
Thus, to determine the $\eta$ and $\eta'$ masses, we need to solve a set
of five coupled differential equations. We find the smoothest numerical
results if we use the equations of motion for $a, \varphi^q,\varphi^s,A_5^q$ and  $A_5^s$,
where
\begin{equation}
\varphi^0 = \sqrt{\frac{2}{3}} \varphi^q - \frac{1}{\sqrt{3} } \varphi^s
\quad\quad\quad
\varphi^8 = \frac{1}{\sqrt{3}} \varphi^q + \sqrt{\frac{2}{3} } \varphi^s
\end{equation}
 The equations are
\begin{equation}
  \partial_z \frac{C^2}{z^3} \partial_z a - \frac{C^2}{z^3} m^2 (
  \sqrt{\frac{2}{3}} \varphi^q - \frac{1}{\sqrt{3}} \varphi^s - a) + 
\frac{\kappa}{z^5}
  v_q^2 v_s ( \sqrt{\frac{2}{3}} \eta^q - \frac{1}{\sqrt{3}} \eta^s - a) = 0
\end{equation}
\begin{equation}
  \partial_z \frac{1}{z} \partial_z \varphi^q - g_5^2 \frac{v^2_q}{z^3}
  (\varphi^q - \eta^q) - g_5^2 \sqrt{\frac{2}{3}} \frac{C^2}{z^3} (
  \sqrt{\frac{2}{3}} \varphi^q - \frac{1}{\sqrt{3}} \varphi^s - a_{}) = 0
\end{equation}
\begin{equation}
  \partial_z \frac{1 + c_4 z^4}{z} \partial_z \varphi^s - g_5^2
  \frac{v^2_s}{z^3} (\varphi^s - \eta^s) + g_5^2 \frac{1}{\sqrt{3}}
  \frac{C^2}{z^3} ( \sqrt{\frac{2}{3}} \varphi^q - \frac{1}{\sqrt{3}}
  \varphi^s - a_{}) = 0
\end{equation}
\begin{equation}
  m^2 z^2 \partial_z \varphi^q - g_5^2 v^2_q \partial_z \eta^q -
  \sqrt{\frac{2}{3}} g_5^2 C^2 \partial_z a = 0
\end{equation}
\begin{equation}
  m^2 z^2 (1 + c_4 z^4) \partial_z \varphi^s - g_5^2 v^2_s \partial_z \eta^s +
  \frac{1}{\sqrt{3}} g_5^2 C^2 \partial_z a = 0
\end{equation}
with $v_q = m_q z + \sigma z^3$, $v_s = m_s z + \sigma z^3$ and $C$ given in
equation \eqref{Ceq}. All the modes have Dirichlet conditions in the UV and Neumann in the
IR. To canonically normalize the fields, we demand
\begin{equation}
\int \mathrm{d} z \left[
\frac{v_q^2}{z^3} \eta^q (\varphi^q-\eta^q )
+\frac{v_s^2}{z^3} \eta^s (\varphi^s-\eta^s )
+\frac{C^2}{z^3} a (\varphi^0-a )
\right]  
= 1
\end{equation}

The resulting masses are shown as a function of $\kappa$ on the left side of Figure 1. These
curves are convergent, and the asymptotic values for large $\kappa$, as compared to the
experimental central values (in MeV) are
\begin{eqnarray}
  m_{\eta} = 520  &\hspace{1em}& (m_{\eta}^{\mathrm{EXP}} = 549) \\
  m_{\eta'} = 867 &\hspace{1em}& (m_{\eta'}^{\mathrm{EXP}} =  957)
\end{eqnarray}
So we are off by 5\% and 9\% respectively. We can also turn off the anomaly by lowering
$\Lambda_{\mathrm{QCD}}$, as shown on the right in Figure 1.

It is worth emphasizing that taking $\kappa\to\infty$ does {\it not} send $m_{\eta'}\to\infty$. In the chiral
Lagrangian, there is a parameter like $\kappa$ which should be proportional to the
anomaly~\cite{'tHooft:1999jc}, 
and provides a mass term for the $U(1)$ pseudoscalar. In that case, taking $\kappa\to\infty$ {\it does}
decouple the $\eta'$, and the correct $\eta'$ mass can only be reproduced by tuning $\kappa$ against the
other chiral symmetry breaking terms in the Lagrangian. In AdS, we could have simply taken $\kappa=\infty$
to begin with, which would be a simpler model with $\kappa$ is replaced by a boundary condition. However, we choose
to allow $\kappa$ to vary because it gives us an additional handle on the $U(1)$ sector.

\begin{figure}[t]
  \begin{minipage}[]{.45\textwidth}
    \begin{center}  
      \epsfig{file=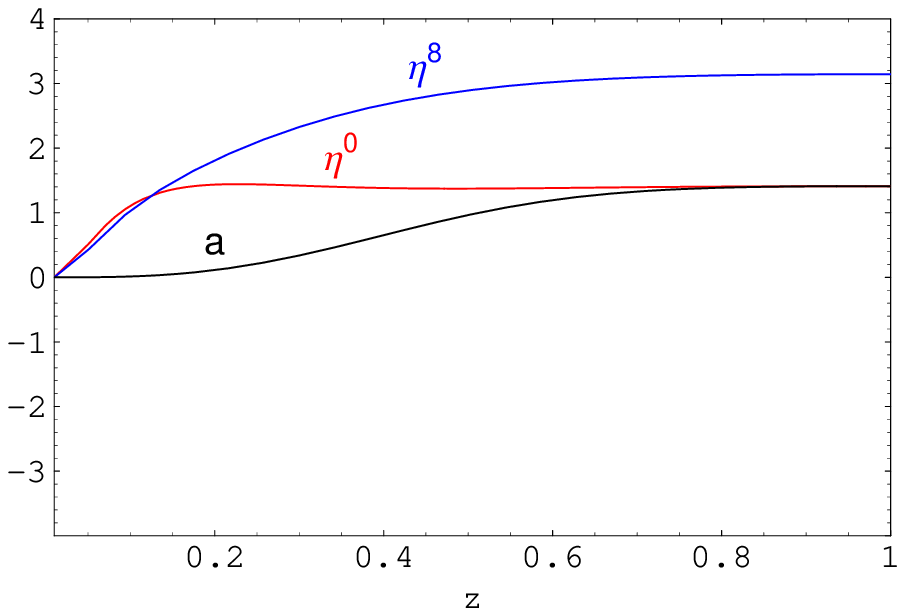, scale=0.75}
    \end{center}
  \end{minipage}
  \hfill
  \begin{minipage}[]{.45\textwidth}
    \begin{center}  
      \epsfig{file=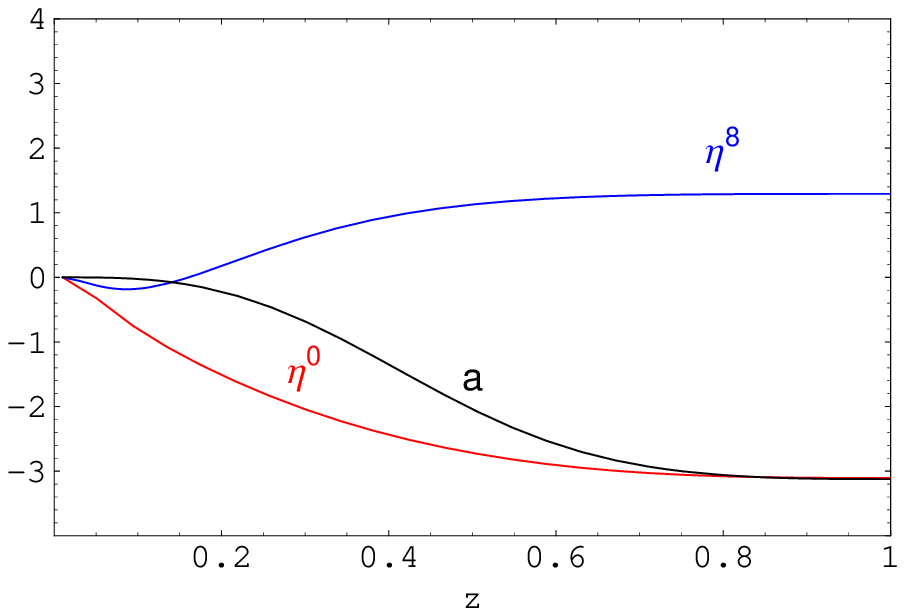, scale=0.75}
    \end{center}
  \end{minipage}
\caption{Profiles of the bulk wavefunctions of the components of $\eta$ (left) and $\eta'$ (right), for
$\kappa=20$. Because of the $z$-dependence, there is no simple mixing-angle interpretation.}
\label{figmodes}
\end{figure}

Next, we calculate the decay constants. There is not a single $f_{\eta}$ and
$f_{\eta'}$. Instead, there is a decay constant for each into the $J^0$ and
$J^8$ currents. 
\begin{eqnarray}
\langle J^\mu_0 | \eta' \rangle &=& i p^\mu f_{\eta 0} \\
\langle J^\mu_8 | \eta' \rangle &=& i p^\mu f_{\eta 8}
\end{eqnarray}
These can be calculated from the wavefunctions directly, using
relations similar to those in~\cite{Erlich:2005qh}. 
For example, we solve the above differential
equations with $m = m_{\eta}$, then evaluate
\begin{eqnarray}
f_{\eta 0} &=& \frac{1}{g_5^2}
\lim_{z \rightarrow  0} \frac{\partial_z \varphi^0}{z} = 17.0 \: \mathrm{MeV} \label{feta0}\\
f_{\eta 8} &=&  \frac{1}{g_5^2}
\lim_{z \rightarrow  0} \frac{\partial_z \varphi^8}{z} = 103 \: \mathrm{MeV}
\end{eqnarray}
Similarly
\begin{eqnarray}
  f_{\eta' 0} &=& 129\: \mathrm{MeV} \\
  f_{\eta' 8} &=& -35.1 \:\mathrm{MeV}
\end{eqnarray}
So qualitatively, the $\eta'$ is more $\eta^0$ and the $\eta$ more $\eta^8$,
as expected. These values
can be compared to decay constants extracted within
chiral perturbation theory~\cite{Feldmann:1999uf}.
It is, however, misleading to represent this mixing in terms of angles
 because the $| \eta^0 \rangle$ and $| \eta^8 \rangle$ components of the
mass eigenstates $| \eta \rangle$ and $| \eta' \rangle$ depend on $z$. This can
be seen from Figure~\ref{figmodes}, which shows the profiles of the AdS
wavefunctions of $\eta$ and $\eta'$.

\begin{figure}[t]
  \begin{minipage}[t]{.45\textwidth}
    \begin{center}  
      \epsfig{file=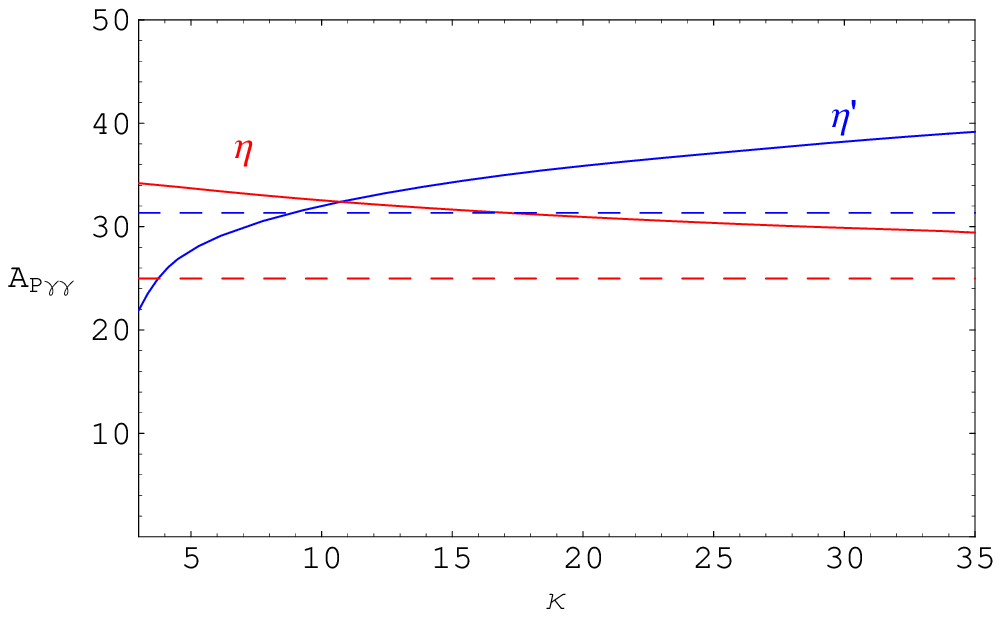, scale=0.75}
    \end{center}
  \end{minipage}
  \hfill
  \begin{minipage}[t]{.45\textwidth}
    \begin{center}  
      \epsfig{file=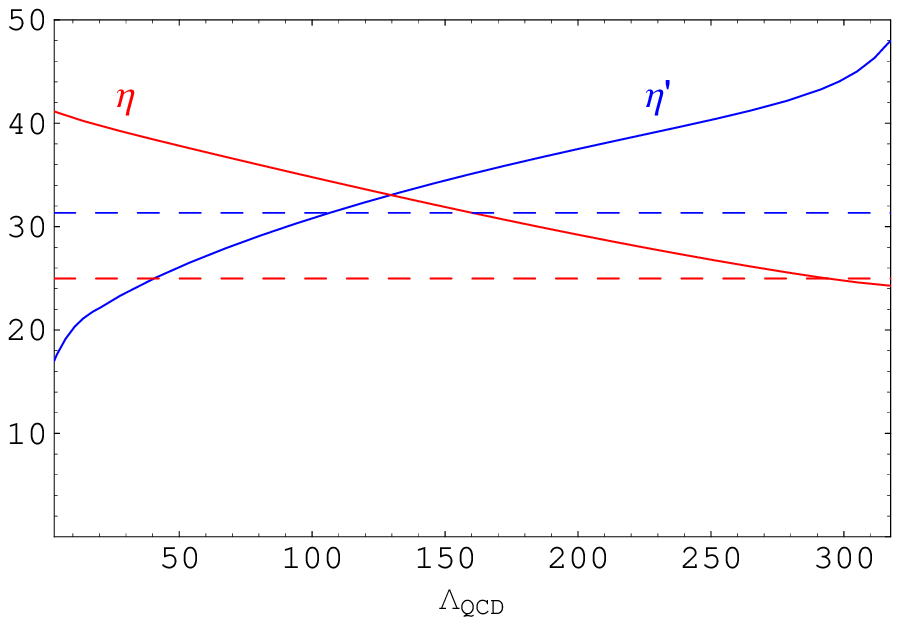, scale=0.75}
    \end{center}
  \end{minipage}
\caption{Decay amplitudes for $\eta$ and $\eta'$ as $\kappa$ and $\Lambda_{\mathrm{QCD}}$ 
are varied. Dashed lines are the experimental values.}
\label{figdecays}
\end{figure}

These decay constants are not directly observable. What is observable are
the neutral pseudoscalar decays $P \rightarrow \gamma \gamma$, which
are mediated by the axial anomaly. Amusingly, the form of this
anomalous interaction in five dimensions was derived long go by Wess, Zumino
and Witten (WZW)~\cite{Wess:1971yu,Witten:1983tw}. 
The bulk Chern-Simons (CS) term relevant for the decay to photons is
\begin{equation}
  \mathcal{L}_{\mathrm{CS}} = \frac{3e^2}{2 \pi^2}
  \varepsilon^{\mathrm{ABCDE}} V_{A B} V_{C D} A_E ^b
  \mathrm{Tr} [Q^2 \tau^b]
\end{equation}
where $Q$ is the generator of electric charge. Here, $V_{\mathrm{AB}}$ are
components of the field strength for the vector gauge field, $V_M (z)$, from
which we want to extract the constant photon zero mode by setting $V_{\mu} (z)
= 1$ (this normalization is consistent with \eqref{u1norm}, 
see~\cite{Katz:2005ir} for more details about the photon). 
In addition to this bulk term, there is a WZW term on the IR boundary at
$z=z_m$
\begin{equation}                                                                                                    
  \mathcal{L}_{\mathrm{WZW}} = 
\frac{3 e^2} {2 \pi^2}  \varepsilon^{\mu\nu\rho\sigma}  
V_{\mu \nu} V_{\rho \sigma} \eta^b \mathrm{Tr} [Q^2 \tau^b]                                                                         
\end{equation}
which absorbs the anomaly.
For constant $V_{\mu}$, with $A_5 = 0$ as usual, the CS term is
a total derivative, and therefore only the boundary WZW term contributes.
Explicitly, the amplitude is
\begin{equation}
  A_{P \gamma \gamma} = \frac{e^2}{4 \pi^2} \left[ \frac{1}{\sqrt{3}} \eta^8
  (z_m) + \frac{4}{\sqrt{6}} \eta^0 (z_m) \right]
\end{equation}
which leads to (as compared to the experimental values extracted from the observed 
decay rates), in units of TeV$^{-1}$
\begin{eqnarray}
  A_{\eta \gamma \gamma} = 24.3,  &\hspace{1em}& (A_{\eta \gamma \gamma}^{\mathrm{EXP}} = 24.9 )\\
  A_{\eta' \gamma \gamma} = 48.1, &\hspace{1em}& (A_{\eta' \gamma \gamma}^{\mathrm{EXP}} = 31.3 )
\end{eqnarray}
These are the asymptotic values at large $\kappa$.
The variation of the decay constants with $\kappa$ 
and $\Lambda_{\mathrm{QCD}}$ is shown in Figure~\ref{figdecays}. 

We can also find the value of $\kappa$ which provides the best fit to the experimental values
of $m_\eta, m_{\eta'}, A_{\eta\gamma\gamma}$ and $A_{\eta'\gamma\gamma}$. This is given by $\kappa=26.1$
with 
$m_\eta = 466\: \mathrm{MeV}, m_{\eta'}=792\: \mathrm{MeV}, A_{\eta\gamma\gamma}=30.2\: \mathrm{TeV}^{-1}$ 
and $A_{\eta'\gamma\gamma}=37.3\: \mathrm{TeV}^{-1}$. The RMS error is 18\%.

\section{Topological Susceptibility, Instantons, and $\bar{\theta}_{}$}
Now, let us turn to the topological susceptibility, $\chi_t$. The standard
argument is that if there are massless quarks in the theory, then $\theta$ is
unphysical, and thus $\chi_t$ must vanish. However, if all quarks are massive,
or there are no quarks at all, then we expect $\chi_t$ to be nonzero. These
facts lead the Witten-Veneziano relation~\cite{Witten:1979vv,Veneziano:1979ec}
for the $\eta'$ mass at large $N_C$
\begin{equation}
\chi_t =  \frac{f_{\eta}^2}{4 N_f} (m_{\eta}^2 + m_{\eta'}^2 - 2 m_K^2)
\end{equation}
This relation, which gives $\chi_t = (171$ MeV)$^4$, is only approximate. It assumes
all the decay constants are equal, that the mesons have no glueball component,
and that $N_C$ is large. Nevertheless, lattice seems to confirm these
approximations~\cite{DelDebbio:2004mc} by producing $\chi_t = (191 \mathrm{MeV})^4$.
With our 5D construction, we can calculate the
topological susceptibility, the meson masses, and the decay constants
directly, and furthermore we can {\it verify} that $\chi_t$ vanishes only
with massless quarks, from which the Witten-Veneziano relation follows.

Recall that
\begin{equation}
  \chi_t = \frac{C^2}{g_a^2} \lim_{z \rightarrow 0} \frac{a \partial_z
  a}{z^3} \label{ateq}
\end{equation}
for a solution with $a (0) = 1$. First, consider the case of pure
gluodynamics. Then there are no $\eta$ or $A_5$ fields, and the equation of motion
at zero momentum is simply
\begin{equation}
  \partial_z \frac{1}{z^3} \partial_z a = 0
\end{equation}
In the absence of a $\kappa$ term, it is simplest to just impose $a (z_m) = 0$
directly. Then the solution is 
$a (z) = 1 - (\frac{z}{ z_m})^4 = 1- \frac{1}{4} (g_a^2/C^2)\chi_t z^4$ 
from \eqref{ateq}.
Now suppose there are quarks. In the limit that the anomaly is weak (for example at large $N_C$), we can
do a perturbation expansion in $C$.
To leading order in $C$, the $\eta-\varphi$ system decouples from the $a$ mode. 
Then from~\eqref{feta0}
we get $\varphi(z) =1+ \frac{1}{2} g_5^2 f_\eta z^2$.
Then the equation of motion~\eqref{A5eqn}, with $Q^2=-m_{\eta}^2$ 
and using
\eqref{Cdef} and \eqref{gadef}, gives
\begin{equation}
\chi_t = \frac{1}{4 N_f} f_{\eta}^2 m_{\eta}^2 
\end{equation}
which matches Witten-Veneziano.

To see that $\chi_t$ vanishes with massless quarks, we no longer assume that $C$ is small.
Then the $\eta$ and $A_5$ equations of motion (at $Q = 0$) are
\begin{equation}
 \partial_z \frac{(m_q z + \sigma z^4)^2}{z^3} \partial_z \eta = 0
\end{equation}
\begin{equation}
(m_q z + \sigma z^4)^2 \partial_z \eta - C^2 \partial_z a = 0
\end{equation}
If $m_q = 0$ then the only solution for $\eta$ satisfying $\eta(0) = 0$ is $\eta(z)
= 0$. Then $a(z)$ must be constant and the topological susceptibility
vanishes. However, as long as $m_q \neq 0$, there is a solution with $\eta \sim
z^2$ and $a \sim z^4$ near $z = 0$. In this case $\chi_t$ is nonzero. In fact,
we can solve the equations exactly for constant $C$ with $\eta (z_m) = a (z_m)$
boundary conditions, giving
\begin{equation}
  \chi_t = \frac{\alpha_s^2}{\pi^4} \frac{m_q z_m (m_q z_m + \sigma z_m^3)}{2 C^2 +
  m_q z_m (m_q z_m + \sigma z_m^3)} z_m^{- 4}
\end{equation}
For $C = 0$ this reduces to the result from pure gluodynamics. If $C \neq 0$,
then we can see directly that $m_q = 0$ forces $\chi_t$ to vanish, as
expected.

Note that we have not used the $\kappa$ term at all to calculate the
topological susceptibility; we have only used the fact that it leads to $a
(z_m) = 0$ in pure gluodynamics, or $a (z_m) = \eta (z_m)$ if quarks are
included. The $\kappa$ term is supposed to represent some non-perturbative
effects which are normally associated with instantons, so it is natural to
ask if we can make the connection more precise. In QCD the one instanton
contribution to the topological susceptibility can be calculated 
explicitly~\cite{Novikov:1979ux,Geshkenbein:1979vb}
\begin{equation}
 \chi_t (Q) = \cdots - \frac{1}{2}
 \int_0^{\infty}\mathrm{d}\rho \frac{{\mathcal{D}}(\rho)}{\rho^5}
\left[  Q^2 \rho^2 \mathcal{K}_2 (Q \rho) \right]^2
\end{equation}
Here, ${\mathcal{D}} (\rho)$ is the dilute-gas instanton density.
For example, for $N_C=3$, 
${\mathcal{D}}(\rho)=(\Lambda_{\mathrm{QCD}} \rho)^{11}$.
The Bessel function ${\mathcal{K}}_2$ appears as the
Fourier transform of $G \tilde{G}$ evaluated on a one-instanton solution. This
expression is divergent due to large instantons, so one
normally cuts off the integral at 
$\rho = \rho_c \sim \Lambda_{\mathrm{QCD}}^{-1}$.

In QCD it is not meaningful to compare the contribution of this specific gauge
configuration to any particular calculation on the AdS side.  This is because the
five-dimensional dual describes only gauge invariant quantities resulting
from integration over all gauge configurations, and it is not clear in which
sense this particular configuration dominates the integral.  
Nevertheless, in truly conformal theories, where the coupling constant is a
marginal parameter, it makes sense to compare non-perturbative contributions to correlators
({\it{i.e.}} in powers of $e^{-1/g^2}$) between the CFT and the five-dimensional
theory.  Of course, the axion would have a similar bulk description in such a 
case, the main difference being the $z$-dependence of the $\kappa$-like term in the
dual to the CFT.  We thus expect that the $\kappa$ term contribution to some
correlation function to have structures similar to those one gets from
integration over instanton size, since the bulk integration over the 
$z$-variable, must ultimately reproduce the same correlation function in the CFT.

To see the similarity to the instanton calculation, let us look at the $G
\tilde{G}$ two-point function from the bulk perspective. We can 
solve for $a$ perturbatively around
the conformal limit. In the conformal approximation, there is no IR brane, and
the axion bulk-to-boundary propagator (with $a(0)=1$) is
\begin{equation}
  z^3 \partial_z \frac{1}{z^3} \partial_z a^{(0)} - Q^2 a^{(0)} = 0 \quad
\Rightarrow \quad
a^{(0)}(z) = \frac{1}{2} z^2 Q^2 {\mathcal{K}}_2(Q z)
\end{equation}
Conformality is broken by the IR brane and by the $\kappa$ term. With these effects,
the equation
of motion becomes
\begin{equation}
C^2  z^3 \partial_z \frac{1}{z^3} \partial_z a - C^2 Q^2 a + \frac{1}{z^5} \kappa(z) a = 0
\end{equation}
where $\kappa(z)$ includes the non-conformal $z$-dependence of the $\kappa$ term. 
Now, we can get a simple expression for the topological susceptibility
by integrating the action by parts on the equations of motion
\begin{equation}
  \int \mathrm{d} z \left[ \frac{C^2}{z^3} (\partial_z a)^2 + \frac{C^2}{z^3} Q^2
  a^2 - \frac{1}{z^5} \kappa(z) a^2 \right] 
=  C^2 \lim_{z \rightarrow 0} \frac{a\partial_z a}{z^3}
= 2 N_f \chi_t (Q)
\end{equation}
If conformal invariance is a good approximation, we can
estimate the effect of conformal symmetry breaking  by evaluating this expression on $a^{(0)}$. We thus
find
\begin{equation}
  \chi_t (Q) = - \frac{1}{8 N_f}\int_0^{z_m} \mathrm{d} z
 \frac{\kappa(z)}{z^5} \left[ Q^2 z^2 {\mathcal{K}}_2 (Q z)  \right]^2
\end{equation}
This has exactly the same form as the instanton contribution. Thus, the scale dependence
of the $\kappa$ term acts just like the instanton density and the IR brane provides
a natural cutoff on the integral over instanton size. 

Finally, let us say a word about the QCD vacuum angle $\theta$. This angle is
intimately tied to the solution of the $U (1)$ problem. The argument, roughly,
is that the topological susceptibility must be nonzero to split the $\eta'$
from the $\eta$ and the $\pi^0$. Since the topological susceptibility is the
second variation of the effective action with respect to $\theta$, there must
be sensitivity to $\theta$ in QCD. Thus the strong CP problem, which is why
the apparent value of $\theta$ is so tiny ($\theta \lesssim 10^{- 9}$), must
be taken seriously.

In AdS, there are three angles, appearing in the $X$ and $Y$ vevs, and in the
$\kappa$ term. We can write $\langle X \rangle = | \langle X \rangle |e^{i
\theta_1}$, \ $\langle Y \rangle = | \langle Y \rangle |e^{i \theta_2}$ and
$\kappa = | \kappa |e^{i \theta_3}$. Since $\theta_1$ and $\theta_2$ come from
vevs, they can be functions of $z$ (as in Eqs.(\ref{xvev},\ref{yvev})), but $\theta_3$, like $\kappa$,
should be a constant. In full generality, $\theta_1$ can have flavor indices
as well. This leads to
\begin{equation}
  \mathcal{L} = \frac{v^2}{2 z^3} [A_5^b + \partial_z (\eta^b - \theta_1^b)]^2 +
  \frac{C^2}{2 z^3} [A_5^0 + \partial_z (a - \theta_2)]^2 + \frac{\kappa}{2 z^5}
  v^{N_f} (a - \eta^0 - \theta_3)^2
\end{equation}
Now, the axial symmetries in AdS are local gauge symmetries, so we can rotate
$\theta_1^b$ and $\theta_2$ into $\eta^b$ and $a$ respectively. This leaves
$\theta_3 = \bar{\theta}$ as the physical vacuum angle. Although the
combination $a - \eta^0$ couples directly to $\bar{\theta}$, it cannot be the
physical axion which solves the strong $\mathrm{CP}$ problem. Even though
$\langle a - \eta^0 \rangle = \bar{\theta}$, $\bar{\theta}$ cannot be
eliminated since it is $a$ and $\eta$ which appear in the rest of the
Lagrangian, not the orthogonal combination $a + \eta^0$. However, there is hope
that since AdS allows a quantitative study of confinement, a strong-dynamics
based solution to the strong CP problem might be realizable.

\section{Summary and conclusions}
We have studied the $U (1)$ problem through and extra-dimensional model
inspired by the AdS/CFT correspondence. This model is built from the bottom
up, by fitting some parameters to perturbative QCD correlation functions
and others to data. All of the parameters in the model can be determined by
the experimental masses of the $\pi^0$, $K^0$ and $\rho$ mesons, and the pion
decay constant $f_{\pi}$. This is only one more experimental value than is
needed to define QCD itself (in QCD, we have the quark masses $m_q$ and $m_s$
and the value of $\Lambda_{\mathrm{QCD}}$). In particular, no strong-dynamics
based observable, such as the topological susceptibility, is needed to study
the $\eta'$. Instead, we only need the coefficient of the anomaly which
is perturbatively calculable and 1-loop finite. The non-perturbative effects
are represented in our model with a $\kappa$ term, on which we have shown the
observables are only weakly dependent. Using this construction, we have
calculated $m_{\eta'} = 867$ MeV, which is 9\% off from experiment. We have
also calculated its decay constants, and its coupling to photons, as well as
the analog quantities for the $\eta$. The best fit for $\kappa$ matches the four
observables to 18\%.

In addition to being quantitatively precise, the extra-dimensional
construction allows for additional qualitative insight into the U(1) problem
and related issues. For example, we have shown how the vanishing of a quark
masses would cause the topologically susceptibility to vanish, independent of
any discussion of the theta angle $\bar{\theta}$ of QCD. From this, the
Witten-Veneziano relations follow. In QCD, it is difficult to study
the contribution of non-perturbative effects, 
because one cannot turn off the anomaly except by taking
$N_C\to\infty$. In the holographic model there are two additional parameters,
$\kappa$ and $\Lambda_{\mathrm QCD}$ which can be separately dialed, giving
us new handles on the anomaly.
We also showed that the non-perturbative
contribution to the topologically susceptibility, which can be represented with
an instanton calculation, has a direct analog in AdS. The same Bessel
functions appear in both cases, and the integral over instanton size is
replaced by an integral over the extra dimension. Instead of having to invoke
a separate cutoff to regulate the IR divergence, we naturally use the same IR
cutoff we would have in a non-anomalous theory. 

This solution to the $U (1)$ problem demonstrates the versatility of
the bottom-up AdS/QCD approach. It also emphasizes that 
AdS is not just a complicated way of phrasing the predictions of chiral
perturbation theory -- the $\eta'$ mass is simply a free parameter in the
chiral Lagrangian. Although our effective description is non-renormalizable,
higher-dimension operators are quantitatively irrelevant for the observables
in question, as is expected from naive dimensional analysis. It is therefore
likely that through further application of the AdS/QCD correspondence,
additional quantitative and qualitative information about the
non-perturbative structure of gauge theories can be derived.

\section*{Acknowledgments}
The authors would like to thank S.~Adler, J.~Maldacena, D.~Son, and E.~Witten for
enlightening discussions.
E.K. was supported in part by the Department of Energy grant
no. DE-FG02-01ER-40676, and by the NSF CAREER grant PHY-0645456.
M.S. was supported in part by the National Science Foundation under grant
NSF-PHY-0401513 and by the Johns Hopkins Theoretical Interdisciplinary Physics and 
Astronomy Ceneter.


\begin{thebibliography} {eta}

\bibitem{Weinberg:1975ui}
  S.~Weinberg,
  Phys.\ Rev.\  D {\bf 11}, 3583 (1975).

\bibitem{'tHooft:1999jc}
  G.~'t Hooft,
  arXiv:hep-th/9903189.

\bibitem{'tHooft:1986nc}
  G.~'t Hooft,
  Phys.\ Rept.\  {\bf 142}, 357 (1986).

\bibitem{Nath:1980nf}
  P.~Nath and R.~Arnowitt,
  Nucl.\ Phys.\  B {\bf 209}, 251 (1982).


\bibitem{Kaiser:1998ds}
  R.~Kaiser and H.~Leutwyler,
  arXiv:hep-ph/9806336.


\bibitem{'tHooft:1976fv}
  G.~'t Hooft,
  Phys.\ Rev.\  D {\bf 14}, 3432 (1976)
  [Erratum-ibid.\  D {\bf 18}, 2199 (1978)].


\bibitem{Novikov:1979ux}
  V.~A.~Novikov, M.~A.~Shifman, A.~I.~Vainshtein and V.~I.~Zakharov,
  Phys.\ Lett.\  B {\bf 86}, 347 (1979)
  [JETP Lett.\  {\bf 29}, 594.1979\ ZFPRA,29,649 (1979\ ZFPRA,29,649-652.1979)].

\bibitem{Aoki:2006xk}
  S.~Aoki {\it et al.}  [JLQCD Collaborations],
  arXiv:hep-lat/0610021.

\bibitem{DelDebbio:2004mc}
  L.~Del Debbio, L.~Giusti and C.~Pica,
  Nucl.\ Phys.\ Proc.\ Suppl.\  {\bf 140}, 603 (2005)
  [arXiv:hep-lat/0409100].

\bibitem{Maldacena:1997re}
  J.~M.~Maldacena,
  Adv.\ Theor.\ Math.\ Phys.\  {\bf 2}, 231 (1998)
  [Int.\ J.\ Theor.\ Phys.\  {\bf 38}, 1113 (1999)]
  [arXiv:hep-th/9711200].

\bibitem{Erlich:2005qh}
  J.~Erlich, E.~Katz, D.~T.~Son and M.~A.~Stephanov,
  Phys.\ Rev.\ Lett.\  {\bf 95}, 261602 (2005)
  [arXiv:hep-ph/0501128].


\bibitem{DaRold:2005zs}
  L.~Da Rold and A.~Pomarol,
  Nucl.\ Phys.\ B {\bf 721}, 79 (2005)
  [arXiv:hep-ph/0501218].


\bibitem{Katz:2005ir}
  E.~Katz, A.~Lewandowski and M.~D.~Schwartz,
  Phys.\ Rev.\  D {\bf 74}, 086004 (2006)
  [arXiv:hep-ph/0510388].


\bibitem{Csaki:1998qr}
  C.~Csaki, H.~Ooguri, Y.~Oz and J.~Terning,
  JHEP {\bf 9901}, 017 (1999)
  [arXiv:hep-th/9806021].

\bibitem{Sakai:2004cn}
  T.~Sakai and S.~Sugimoto,
  Prog.\ Theor.\ Phys.\  {\bf 113}, 843 (2005)
  [arXiv:hep-th/0412141].


\bibitem{Shifman:1978bx}
  M.~A.~Shifman, A.~I.~Vainshtein and V.~I.~Zakharov,
  Nucl.\ Phys.\  B {\bf 147}, 385 (1979).

\bibitem{Reinders:1981ww}
  L.~J.~Reinders, S.~Yazaki and H.~R.~Rubinstein,
  Nucl.\ Phys.\  B {\bf 196}, 125 (1982).

\bibitem{Randall:2001gb}
  L.~Randall and M.~D.~Schwartz,
  JHEP {\bf 0111}, 003 (2001)
  [arXiv:hep-th/0108114].


\bibitem{Feldmann:1999uf}
  T.~Feldmann,
  Int.\ J.\ Mod.\ Phys.\  A {\bf 15}, 159 (2000)
  [arXiv:hep-ph/9907491].

\bibitem{Wess:1971yu}
  J.~Wess and B.~Zumino,
  Phys.\ Lett.\  B {\bf 37}, 95 (1971).

\bibitem{Witten:1983tw}
  E.~Witten,
  Nucl.\ Phys.\  B {\bf 223}, 422 (1983).

\bibitem{Witten:1979vv}
  E.~Witten,
  Nucl.\ Phys.\  B {\bf 156}, 269 (1979).

\bibitem{Veneziano:1979ec}
  G.~Veneziano,
  Nucl.\ Phys.\  B {\bf 159}, 213 (1979).

\bibitem{Geshkenbein:1979vb}
  B.~V.~Geshkenbein and B.~L.~Ioffe,
  Nucl.\ Phys.\  B {\bf 166}, 340 (1980).

\end{thebibliography}
\end{document}